\documentclass[seceq]{ptptex}

\usepackage{graphicx}

\title{
Revenue Prediction of Local Event using Mathematical Model of Hit Phenomena%
}

\author{
Akira  \textsc{Ishii}, %
Takehiro  \textsc{Matsumoto}, %
Shinji  \textsc{Miki}, %
}

\inst{
Department of Applied Mathematics and Physics, Tottori University, Koyama, Tottori 680-8552, Japan

}

\date{}

\abst{
Theoretical approach to investigate human-human interaction in society performed using a many-body theory including human-human interaction. The advertisement is treated as an external force. The word of mouth (WOM) effect is included as a two-body interaction between humans. The rumor effect is included as a three-body interaction between humans. The parameters to define the strength of human interactions are assumed to be constant values. The calculated result explained well the two local events "Mizuki-Shigeru Road in Sakaiminato" and  "the sculpture festival at Tottori"  in Japan.  

}

\begin{document}

\maketitle

\section{Introduction}

  The human interaction in real society can be considered as many body theory of human. Especially after the population of the social network systems (SNS) like blogs, Twitter, Facebook, Google+ or other similar services in the world, the interactions between each accounts can be stocked as digital data. Though the SNS society is not equal to the real society, we can assume that the communication in SNS society is very similar to that in the real society. Thus, we can use huge amount of stock of digital data of human communication as observation data of the real society.\cite{Allsop2007}, \cite{Kostka2008},\cite{Bskshy2011},\cite{Jansen2009}
 Using this observation, we can apply the method of statistical mechanics to social sciences. Since the word-of-mouth (WOM) has bee pointed out to be very significant for example in marketing science\cite{Brown1987},\cite{Murray1991}, \cite{Banerjee1992},\cite{Taylor2003}, such analysis and prediction of the digital WOM in the sense of statistical physics will be important today.
  
    In the previous paper, as an applied field of the statistical mechanics of human dynamics, A,Ishii, one of the author, focused his attention to the motion picture entertainment of the Japanse market.  In this study, we consider the hit phenomena of movies with both experimental and theoretical ways. For experimental viewpoint, we observe daily revenue data and daily blog-writing data for many movies. We also obtain the advertisement cost for the movies. For theoretical viewpoint, we present here a mathematical model for hit phenomena as non-equilibrium, nonlinear and dynamical phenomena. Comparing the simulated revenue with the observed revenue for the movies checks the model. The calculations for movie presented in the previous works\cite{ishii2005},\cite{ishii2007},\cite{Ishii2010},\cite{Yoshida2010},\cite{Ishii2011a},\cite{Ishii2011b} that our calculation explain the observed revenue or the observed daily number of weblog posting very well. 
    
    In this paper, we apply the theory to the local event in Japan. The purpose to apply to the local event is that our theory can be applied not only the motion picture entertainments but many other economic activities in the real society.

\section{Theory}

\subsection{Purchase intention for individual person}

Based on the observation of the hit phenomena in Japanes market, we present a theory to explain and predict hit phenomena. First, instead of the number of audience $N(t)$, we introduce here the integrated purchase intention of individual customer, $J_i(t)$ defined as follows,

\begin{equation} 
\label{eq:eq3}
N(t) = \sum_i J_i(t)
\end{equation}

here the suffix j corresponds to individual person who has attention to go to the local event he/she concern. 

The daily purchase intention is defined from $J_i(t)$ as follows, 
\begin{equation}
\frac{dJ_i(t)}{dt} = I_i(t)
\end{equation}

The number of integrated audience or incoming people to the event can be calculated using the purchase intention as follows,

\begin{equation}
N(t) = \int_0^t \sum_i I_i(\tau) d\tau
\end{equation}

  Since the purchase intention of the individual customer increase due to both the advertisement and the communication with other persons, we construct a mathematical model for hit phenomena as the following equation. 

\begin{equation}
\frac{dI_i(t)}{dt} =  advertisement(t) + communication(t)
\end{equation}

\subsection{advertisement}

Advertisement is the very important factor to increase the purchase intention of the customer in the market. Usually, the advertisement campaign is done at TV, newspaper and other medias. We consider the advertisement effect as an external force to the purchase intention as follows,

\begin{equation}
\label{eq:eq12}
\frac{dI_i(t)}{dt} = A(t) + \sum_j D_{ij} I_j(t) * \sum_j \sum_k P_{ijk} I_j(t) I_k(t)
\end{equation}

where $D_{ij}$ is the factor for the direct communication and $P_{ijk}$ is the factor for the indirect communication. Because of the term of the indirect communication, this equation is a nonlinear equation.

\subsection{Mean field approximation}

To solve the equation (\ref{eq:eq12}), we introduce here the mean field approximation for simplicity. Namely, we assume that the every person moves equally so that we can introduce the averaged value of the individual purchase intention. 

\begin{equation}
I = \frac{1}{N_p} \sum_j I_j(t)
\end{equation}

where introducing the number of potential audience $N_p$.  We obtain the direct communication term from the person who do not watch the movie as follows,

\begin{equation}
\sum_j D_{ij} I_j(t) = N_p \frac{1}{N_p} \sum_j D_{ij} I_j(t)  \Rightarrow \frac{N_p -N(t)}{N_p} (N_p-N(t))D^{nn}I
\end{equation}

where $D^{nn}$ is the factor of the direct communication between the persons who do not watch the movie at the time $t$. 

Similarly, we obtain the indirect communication term due to the communication between the person who do not watch the movie at the time $t$, 

\begin{eqnarray}
\sum_j \sum_k P_{ijk} I_j(t) I_k(t) = N_p \frac{1}{N_p} \sum_j N_p \frac{1}{N_p} \sum_k  \nonumber  \\
\Rightarrow (\frac{N_p-N(t)}{N_p})^3 N_p^2 P^{nn} I^2 = \frac{(N_p-N(t))^3}{N_p} P^{nn} I^2
\end{eqnarray}

where $P^{nn}$ is the factor of the indirect communication between the persons who do not watch the movie at the time $t$.

  For the direct communication between the watched person and the unwatched person can be written as follows,
  
\begin{equation}
\sum_j D_{ij} = N_p \frac{1}{N_p} \sum_j D_{ij} I_j(t) \Rightarrow \frac{N(t)}{N_p} (N_p-N(t)) D^{ny} I
\end{equation}  

where $D^{ny}$ is the factor of the direct communication between the watched person and the unwatched person. For the indirect communication, we obtain more two terms corresponding to the indirect communication due to the communication between the watched persons and that between the watched person and the unwatched person as follows,

\begin{equation}
\frac{(N(t))^2 (N_p-N(t))}{N_p} P^{yy} I^2 + \frac{N(t) (N_p-N(t))^2}{N_p} P^{ny} I^2
\end{equation}

where $P^{yy}$ is the factor of the indirect communication between the watched persons and $P^{ny}$ is the factor of the indirect communication due to the communication between the watched person and the unwatched person at the time $t$. 

  Finally, we obtain the equation of the mathematical model for hit phenomena within the mean field approximation as follows,

\begin{eqnarray}
\label{eq:eq22}
\frac{I(t)}{dt} &=& (A(t) + \frac{N(t)}{N_p} (N_p-N(t)) D^{ny} I \nonumber \\   
&& + \frac{(N(t))^2 (N_p-N(t))}{N_p} P^{yy} I^2 + \frac{N(t) (N_p-N(t))^2}{N_p} P^{ny} I^2 )bf(t)\nonumber \\
&&
+ \frac{N_p -N(t)}{N_p} (N_p-N(t))D^{nn}I +  \frac{(N_p-N(t))^3}{N_p} P^{nn} I^2
\end{eqnarray}

where

\begin{equation}
\label{eq:eq23}
N(t) = N_p \int_0^t I(\tau) d\tau
\end{equation}

and $f(t)$ means the rush effect or the increasing number of incoming people toward the last day of the event. $b$ is the parameter to represent the difference of the incoming between the weekdays and weekend.

\begin{equation}
\label{eq:eq23a}
f(t) = c(t_{end}-t)^{-4}
\end{equation}

The power -4 is decided by adjusting our calculation to the observed data of the rush effect of the world sand sculpture festival 2009 in Tottori Japan.

The equations  (\ref{eq:eq22}) with (\ref{eq:eq23}) is based on the equation we presented in the previous work for the motion picture entertainment market \cite{Ishii2010},\cite{Yoshida2010},\cite{Ishii2011a}. This equation is, in principle, also derived in our recent paper \cite{Ishii2011b} using the stochastic processes. Thus, the equation (\ref{eq:eq22}) with (\ref{eq:eq23}) is the nonlinear integrodifferential equation. However, since the handling data is daily, the time difference is one day, we can solve the equation numerically as a difference equation.

\section{Results}

\subsection{Target Local Event, Mizuki-Shigeru Road}

The location of Tottori prefecture is far from Tokyo and the Sakaiminato city stays as countryside. Mr.Mizuki, Shigeru, who was born in Sakaiminato-city, Tottori of Japan, produced many pieces of work based on the motif of Yokais(Demons) such as "Ge Ge Ge no Kitaro." He was awarded Shiju Hosho Decoration in 1991, Minister of Edcation Award of Japan in 1996 and Kyokujitu Sho Decoration in 2003, for his excellent achievements and cultural contribution.  In the Sakaiminato city, there is Mizuki-Shigeru Road where 134 yokai characters were sterically casted into detailed bronze statues. Anyone can see and touch the statues. Each unit of the black granite base and the statue is highly appraised as a fine art. Yokais attract a number of visitors and keep the town active.

Tourism Section of the Trading and Tourism Divison of the Industry Environment Department of Sakaiminato City keep the daily advertisement costs and the daily number of incoming guest to the Mizuki-Shigaru Road for recent 10 years. Therefore, we use the data  for the application of our theory to the real local event. The daily number of weblogs posting for the Mizuki-Shigeru Road was collected by using the site service called "Kizasi".  We found the number of blog posting is very similar to the revenue of corresponding movie. The number of blog-writing entry can be used as quasi-revenue.\cite{Ishii2010},\cite{Yoshida2010},\cite{Ishii2011a},\cite{Ishii2011b}  
 
First, we show daily positive, middle and negative posting of weblog for Sakaiminato Mizuki-Shigeru Road and Daily Advertisement cost for the Mizuki-Shigeru Road in April-July, 2008 in fig.1. The judgement of "positive" means that the author want to visit Sakaiminato to enjoy Mizuki-Shigeru Road. The judgement of "negative" means the opposite meaning. 

In fig.1, It should be noticed that there are holiday seasons in Japan from the April 29 to May 5. Thus, we found that the large peak of the positive weblog around the beginning of May. In the figure we found that the dominant postings are positive posting. Thus, we can use the daily number of weblog as the purchase intention. 

\begin{figure}[htbp]
\begin{center}
\includegraphics[scale=.8]{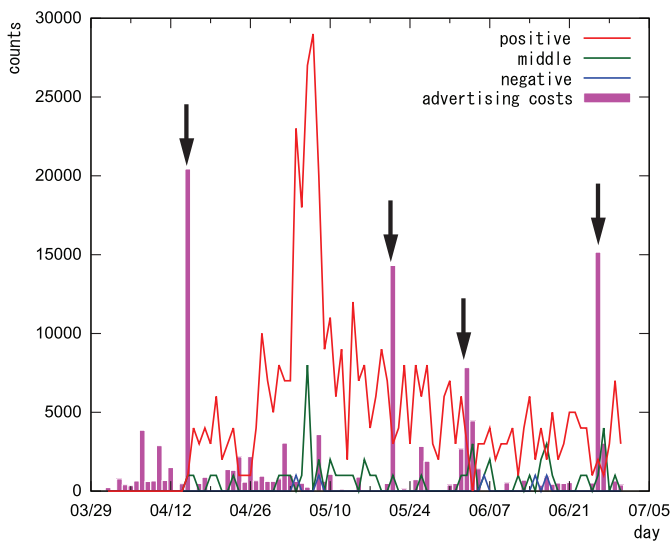}
\caption{Daily positive, middle and negative posting of weblog for Sakaiminato Mizuki-Shigeru Road and Daily Advertisement cost for the Mizuki-Shigeru Road in April-July, 2008.}
\label{fig:fig1}
\end{center}
\end{figure} 

\subsection{Decline of advertisement effect}

Moreover, we found that the peak of the daily advertisement cost and the peak of the positive  blog is shifted one day or two days. The arrows in fig.1 indicate the advertisement costs we concern. The peaks of the blog posting are delayed. From these data for 2008 and 2009, we found that the decline rate of the advertisement effect to be nearly 0.4 [1/days] where effective advertisement effect in eq. (\ref{eq:eq22}) is defined as

\begin{equation}
\label{eq:eq24}
A(t) = c \int^{t}_{-\infty} A(\tau) e^{a \tau} d \tau .
\end{equation}

\subsection{Strength of advertisement effect}

From fig.1, we can also find the strength of advertisement effect for the Mizuki-Shigeru Road. The effect of the strength of the advertisement effect is defined as the increasing of the number of incoming or increasing of the revenue of the Mizuki-Shigeru Road. 

\subsection{Result}

Using the decline factor of the advertisement effect and the strength of the advertisement effect, we calculate the purchase intention for the Mizuki-Shigeru Road of 2008. The parameters of the direct and indirect communications are adjusted to explain the observed daily weblog posting data. The calculated results are shown in fig.2. The numerical value of parameters we adjusted is shown in Table 1. 
 
\begin{figure}[htbp]
\begin{center}
\includegraphics[scale=.8]{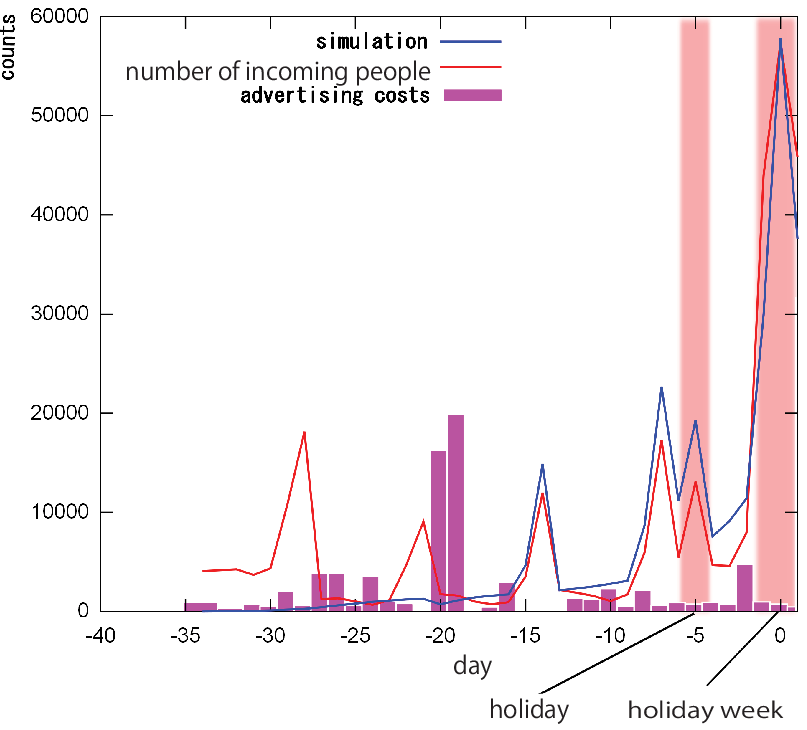}
\caption{Theoretical calculation and the daily positive  posting of weblog for Sakaiminato Mizuki-Shigeru Road in April-July, 2008.}
\label{fig:fig2}
\end{center}
\end{figure} 

\begin{table}[htb]
@\caption{Table of parameters for the Mizuki-Shigeru Road. $N_{p}$ is the number of potential customers. $C_{adv}$ is the strength of advertisement. a is the decline factor of advertisement.. $D_{nn}$ is the direct communication factor. $P_{nn}$ is the indirect communication factor.}
  \begin{center}
   \begin{tabular}{lcrr}
    $N_{p}$ & 500000  \\
    $C_{adv}$ & 0.03  \\
    a & 0.4  \\
    $N_{p}D_{nn}$ & 0.005  \\
    $N_{p}^{2}P_{nn}$ & 0.000013  
  \end{tabular}
  \end{center}
\end{table}

In figure 3, we show the similar calculation for the World Sand Sculpture Festival 2009 held at Tottori-City in Tottori, Japan. The rush effect is very important for this event and the rush power is adjusted to be -4. Compare to the daily positive posting of weblog, we found that the agreement is very well. 
 
\begin{figure}[htbp]
\begin{center}
\includegraphics[scale=.8]{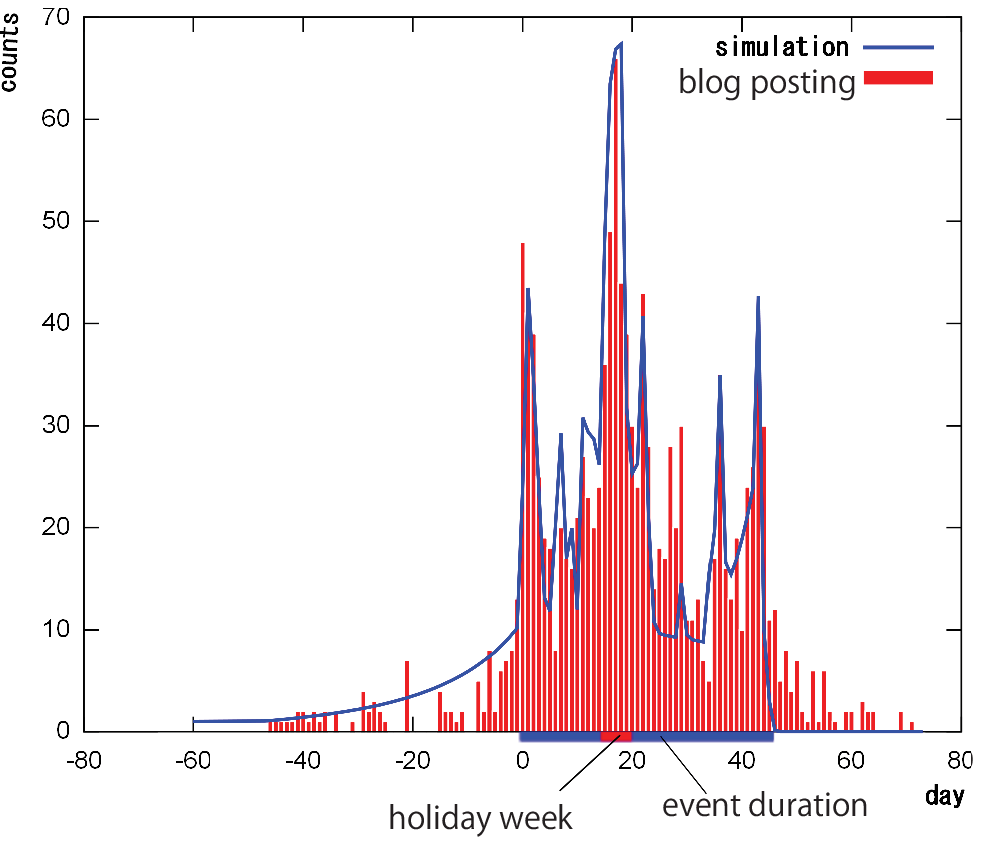}
\caption{Theoretical calculation and the daily positive  posting of weblog for  the World Sand Sculpture Festival 2009 held at Tottori-City in Tottori, Japan..}
\label{fig:fig3}
\end{center}
\end{figure} 
 
\begin{table}[htb]
@\caption{Table of parameters for the the World Sand Sculpture Festival 2009 held at Tottori-City in Tottori, Japan.. $N_{p}$ is the number of potential customers. $C_{adv}$ is the strength of advertisement. a is the decline factor of advertisement.. $D_{nn}$ is the direct communication factor. $P_{nn}$ is the indirect communication factor. "before" and "after" indicate the parameters before the event open and after the event open, respectively}
  \begin{center}
   \begin{tabular}{lcrr}
    $N_{p}$ & 300000  \\
    $C_{adv}$ & 0.0013  \\
    a & 0.01  \\
    $N_{p}D_{nn}(before)$ & 0.04  \\
    $N_{p}D_{nn}(after)$ & 0.037  \\
    $N_{p}D_{ny}$ & 0.00003  \\
    $N_{p}^{2}P_{nn}(before)$ & 0.003  \\
    $N_{p}^{2}P_{nn}(after)$ & 0.002  \\
    $N_{p}^{2}P_{ny}$ & 0.0001  \\
    $N_{p}^{2}P_{yy}$ & 0.00005  
  \end{tabular}
  \end{center}
\end{table}

\section{Disucssion}

From fig.2 and fig.3, we found that our simulation can reproduce the daily counts of incoming 
persons to the local event. Since our theory is very general, it means that our theory can be applied not only to the motion picture entertainment but also to many other market including local events. Therefore, our theory eq. (\ref{eq:eq22}) can be used to investigate human-human interaction especially for entertainment. The assumption to construct our theory is that the price of the target is not expensive so that consumer can be affected by advertisement.  Thus, to produce local events, the producer should use our theory to estimate the rough number of incoming people from the time distribution of advertisement costs. 

The significant point to apply our theory to the Mizuki-Shigeru Road and the world sand sculpture festival 2009 is that the both event counts the number of daily incoming guests exactly using photosensors. Because of this data, we decided to apply our theory to the two events as test cases. Thus, it is very important to check the daily revenue and the daily advertisement costs to predict the near future revenue.

\section{Conclusion}

  The mathematical model for hit phenomena is presented including the advertisement cost and the communication effect. In the communication effect, we include both the direct communication and the indirect communication. The results calculated with the model can predict the revenue of corresponding local events very well. The conclusion presented in this paper is very useful also in marketing science.

\section*{Acknowledgements}
The advertisement cost data and the revenue data are presented by Tourism Section of the Trading and Tourism Divison of the Industry Environment Department of Sakaiminato City. The research is partially supported by the Tottori university research project for rural sustainability.
The authors thank the Yukawa Institute for Theoretical Physics at Kyoto University. Discussions during the YITP workshop YITP-W-11-04 on "econophysics 2011" were useful to complete this work.

%

\end{document}